# Self-consistent magnetic fields and currents

Yurii A. Spirichev

The State Atomic Energy Corporation ROSATOM, "Research and Design Institute
of Radio-Electronic Engineering" Zarechny, Penza region, Russia
E-mail: yurii.spirichev@mail.ru
25.12.2022 г.

**Abstract**

The possibility of the existence of quasi-stationary electromagnetic fields in plasma supported by their own self-consistent current follows from Maxwell's equations with field sources. These equations also give rise to a wave equation for the density of a self-consistent current and a nonlinear kinematic equation for the velocity of electric charges, from which follows the possibility of excitation of a "magnetic dynamo" and localized rotating plasma structures having a magnetic field. An analytical boundary condition for the excitation or attenuation of turbulence and "magnetic dynamo" is obtained.

**Contents**
1. Introduction
2. Wave equation of self-consistent current
3. Nonlinear kinematic equation of charge motion
4. Conclusion
References

### 1. Introduction

In plasma physics, quasi-stationary magnetic fields supported by its own currents are known [1]. In astrophysics and magnetic hydrodynamics, this is the effect of a "magnetic dynamo", manifested in cosmic plasma [2-6] in the form of self-excited and long-existing magnetic fields. In natural terrestrial conditions, the plasma phenomenon of "ball lightning" is known, but still poorly studied [7-11]. "Ball lightning", according to the observations of eyewitnesses, has a strong magnetic field, therefore, it is a localized structure with its own electric currents flowing in it. Laboratory rotating plasma formations with their own magnetic field are known, which are formed under certain conditions when a laser pulse is applied to a metal target or electrical explosions of metal wires [12-14]. Such electric currents and the electromagnetic fields supported by them in plasma are called self-consistent. The review [1] notes that numerous attempts at kinetic description of these plasma effects do not give a clear understanding of the possible types of self-consistent magnetic structures and their characteristic properties. It is also noted that the methods of magnetic hydrodynamics still do not allow us to see a clear picture of the mechanisms of collective interaction of charged plasma particles and the self-consistent magnetic field created by them. This problem is also complicated by the fact that self-consistent electromagnetic processes in plasma are inextricably linked with its turbulence [3-6], which itself is an important problem of plasma physics.

The magnetic field is excited by electric currents, therefore, all the diverse forms of self-consistent quasi-stationary magnetic structures in plasma are associated with their corresponding structures of self-consistent currents, while these structures are inseparable from each other. The general property of the formation of self-consistent structures of the magnetic field and electric current should be implicitly contained in Maxwell's equations describing the relationship of the electromagnetic field with its sources - charges and currents:

$$\nabla \times \mathbf{H} = \mathbf{J} + \partial_t \mathbf{D} \qquad (1)$$

$$\nabla \times \mathbf{E} = -\partial_t \mathbf{B} \qquad (2)$$

$$\nabla \cdot \mathbf{B} = 0 \qquad (3)$$

$$\nabla \cdot \mathbf{D} = \rho \qquad (4)$$

Where $\mathbf{H}$ – the magnetic field strength, E is the electric field strength, $\mathbf{D}$ – electric induction, $\mathbf{B}$ – magnetic induction, $\mathbf{J} = \rho \cdot \mathbf{V}$ – current density, $\rho$ – charge density, $\mathbf{V}$ – charge velocity. For vacuum $\mathbf{B} = \mu_0 \mathbf{H}$ and $\mathbf{D} = \varepsilon_0 \mathbf{E}$. Indeed, from equation (1) follows the law of conservation of charge in the form of a continuity equation for the current density, which obeys all the movements of electric charges:

$$\nabla \cdot \mathbf{J} + \partial_t \rho = 0 \qquad (5)$$

However, this equation alone is not enough to reveal explicitly the general properties of the formation of magnetic and current structures. In Maxwell's equations, electric charges have no mass, so in plasma physics, Maxwell's equations are supplemented with equations of motion for the mass density of electric charges or Newton's second law. The introduction of mass for electric charges means the introduction of mechanical inertia into the equations of their motion, which does not change the electromagnetic physics of their interaction, since the gravitational interaction between charges is not taken into account because of its smallness. Thus, in order to explicitly identify the general rule for the formation of self-consistent magnetic structures and currents, it is sufficient to consider Maxwell's equations and the continuity equation of the current density, since it is the current that is the source of the magnetic field.

The purpose of this article is to derive from Maxwell's equations the equations of self-consistent motion of free electric charges, i.e. electric currents, and to describe the general laws of formation of self-consistent magnetic structures explicitly.

**2. Wave equation of self-consistent current**

From Maxwell's equations follows an inhomogeneous wave equation for an electric field [15] describing electromagnetic waves:

$$\Delta \mathbf{E} - \partial_{tt}\mathbf{E}/c^2 = \mu_0 \partial_t \mathbf{J} + \nabla \rho / \varepsilon_0 \qquad (6)$$

The right-hand side of this equation describes the source of electromagnetic waves in the form of a changing density of charges and currents. Obviously, the continuity Eq. (5) must also be observed. Consider Eq. (6) together with the continuity equation of the current density (5). Take the time derivative of both parts of Eq. (6) and, replacing the time derivative of the charge density from Eq. (5) in the right part, we obtain the equation:

$$\varepsilon_0(\Delta(\partial_t \mathbf{E}) - \partial_{tt}(\partial_t \mathbf{E})/c^2) = \partial_{tt}\mathbf{J}/c^2 - \nabla(\nabla \cdot \mathbf{J}) \qquad (7)$$

The left part of this equation describes the displacement current waves. The right part also has a wave character and describes current density waves, but its spatial derivatives are more complex than those of the classical the D'alembert wave equation. The wave Eq. (7) follows from Maxwell's equations and has a degree of generalization of the description of the motion of electric charges no less than the Maxwell equations themselves for the electromagnetic field. Using the well-known vector identity, we decompose the last term of the right part:

$$\varepsilon_0(\Delta(\partial_t \mathbf{E}) - \partial_{tt}(\partial_t \mathbf{E})/c^2) = \partial_{tt}\mathbf{J}/c^2 - \Delta \mathbf{J} - \nabla \times \nabla \times \mathbf{J} \qquad (8)$$

Now the right part of Eq. (8) contains the classical wave component and the vortex component of the current in the form of a double rotor of current density. Since the right and left sides of this equation are equal, the interaction of an electromagnetic wave with a conducting medium excites current density waves in it, described by the right side of Eq. (8). In magnetic hydrodynamics, displacement currents are usually neglected because of their smallness, compared with conduction currents [16]. To simplify the analysis of the equation, we do the same, equating the left part of Eq. (8) with the displacement currents to zero, we obtain a homogeneous wave equation for the current density:

$$\partial_{tt}\mathbf{J}/c^2 - \Delta \mathbf{J} - \nabla \times \nabla \times \mathbf{J} = 0 \qquad (9)$$

The wave Eq. (9) differs from the canonical form of the D'alembert wave equation by a vortex term $\nabla \times \nabla \times \mathbf{J}$ indicating that current density waves have a vortex component. Let's look at this term in more detail. Replacing the current density in it with its expression $\mathbf{J} = \rho \cdot \mathbf{V}$, we get an expression for the velocity of the charges:

$$\nabla \times \nabla \times (\rho \cdot \mathbf{V}) = \nabla \times (\rho \cdot \nabla \times \mathbf{V} - \mathbf{V} \times \nabla \rho) = 2\rho \cdot \nabla \times \mathbf{\Omega} - 2\mathbf{\Omega} \times \nabla \rho - \nabla \times (\mathbf{V} \times \nabla \rho) \qquad (10)$$

Where $\mathbf{\Omega}$ is the angular velocity of the charges. The first term has the form of the product of the vector of the rotor by the charge density. In accordance with Stokes' theorem, the flow of the rotor vector through the surface is equal to its circulation along a closed contour on which this surface rests, therefore, the flow of the angular velocity vector $\mathbf{\Omega}$ has a circulation along a closed contour. But the angular velocity vector $\mathbf{\Omega}$ itself represents the rotor of the velocity vector $\mathbf{V}$, hence, the density of the velocity vector $\mathbf{V}$ also has a closed loop circulation. Consequently, the total circulation of the density of the velocity vector $\mathbf{V}$ occurs along a closed spiral contour. Thus, the current density waves described by Eq. (9) have a closed spiral-vortex component, which is a toroidal current. In the works [2,3] considering the "magnetic dynamo" it is indicated that its existence requires the existence of toroidal currents that create a poloidal magnetic field. Thus, the Eq. (10) included in the wave Eq. (9) is necessary for the existence of a "magnetic dynamo".

The motion of plasma in a turbulent state is a set of vortex waves [10]. Then the presence of the vortex term (10) in Eq. (9) can be associated with the excitation of plasma turbulence. "Magnetic dynamo", like plasma turbulence, can spontaneously increase in time. From Eq. (9) we can draw a conclusion on this issue. Replacing the current density in Eq. (9) with that taken from Maxwell's Eq. (1), we obtain the equation for magnetic induction **B** and displacement current:

$$\partial_{tt}(\nabla \times \mathbf{B} - \partial_t \mathbf{E}/c^2)/c^2 + \partial_t \nabla \rho / c^2 \varepsilon_0 = 0 \tag{11}$$

Neglecting the displacement current, as is customary in magnetic hydrodynamics [16], we obtain the equation:

$$\partial_{tt}(\nabla \times \mathbf{B}) + \mu_0 \cdot \partial_t \nabla \rho = 0 \tag{12}$$

Integrating it twice in time, we get:

$$\nabla \times \mathbf{B} = -t \cdot \mu_0 \nabla \rho + t \cdot f(\mathbf{r}) \tag{13}$$

Where $f(\mathbf{r})$ is an arbitrary function **r**. It follows from this expression that the rotor increases linearly in time and is directly proportional to the charge density gradient. Consequently, the increasing magnetic field can be excited by the potential field of the charge density gradient. Thus, equation (9) for the current density implies the possibility of magnetic induction an increase in time, which can be qualitatively interpreted as the excitation of a "magnetic dynamo" or plasma turbulence.

**4 Nonlinear kinematic equation of motion of charges**

The left part of the wave Eq. (6) has nontrivial solutions when its right part is equal to zero. For this case, we equate the right side of Eq. (6) to zero and consider it together with the continuity equation of the current density (5) as a system of equations. Substituting an expression for the current density $\mathbf{J} = \rho \cdot \mathbf{V}$ into these equations, and making the necessary transformations, we obtain a system of equations for the density of charges and the speed of their self-consistent motion:

$$\partial_t \rho + \mathbf{V} \cdot \nabla \rho + \rho \cdot \nabla \cdot \mathbf{V} = 0 \tag{14}$$

$$\nabla \rho + \frac{1}{c^2}(\mathbf{V} \cdot \partial_t \rho + \rho \cdot \partial_t \mathbf{V}) = 0 \tag{15}$$

Having solved this equation with respect to velocity, we obtain a nonlinear equation for the velocity of the charges (the derivation of the equation is given in Appendix A):

$$\nabla[(\frac{1}{c^2} \cdot \mathbf{V} \cdot \partial_t \mathbf{V} - \nabla \cdot \mathbf{V})/(1 - \frac{V^2}{c^2})] - \frac{1}{c^2} \cdot \partial_t[(\mathbf{V} \cdot \nabla \cdot \mathbf{V} - \partial_t \mathbf{V})/(1 - \frac{V^2}{c^2})] = 0 \tag{16}$$

This equation is invariant for relativistic velocities of charge motion, since it is derived from invariant relativistic Maxwell equations. For non-relativistic charge velocities, this equation is reduced to a simpler form:

$$\frac{1}{c^2}\partial_{tt}\mathbf{V} - \nabla(\nabla \cdot \mathbf{V}) + \partial_t(\nabla \mathbf{V}^2/2 - \mathbf{V}\cdot\nabla\cdot\mathbf{V})/c^2 = 0 \qquad (17)$$

This nonlinear equation consists of a linear wave part, in form completely analogous to the wave Eq. (9):

$$\frac{1}{c^2}\partial_{tt}\mathbf{V} - \nabla(\nabla \cdot \mathbf{V}) \qquad (18)$$

and a nonlinear vector:

$$\partial_t(\nabla \mathbf{V}^2/2 - \mathbf{V}\cdot\nabla\cdot\mathbf{V})/c^2 \qquad (19)$$

The linear wave part (18) follows from Eq. (9) if the charge density in it is assumed to be equal to a constant value. Consequently, the nonlinear vector (19) is completely determined by the distribution and change in the density of charges as they move in the plasma. In the theory of hyperbolic partial differential equations, for example in the "telegraphic equation" [17], a similar term with the first derivative in time is associated with losses in the physical system and process attenuation. As can be seen from expression (19), the sign of this term may vary depending on the values of its components. This suggests that it can play a dissipative role, as in the "telegraph equation", or, conversely, amplifying the wave process. To enhance the wave process described by Eq. (17), expression (19) must have a negative sign. The first potential term in expression (19) is related to the potential energy of the charges, and the second term is related to their kinetic energy. If the difference of these vectors is negative, then the expression (19) changes the sign to the opposite and the process is intensified. This means that the process increases with an increase in the kinetic energy of the charges, for example, associated with an increase in plasma temperature. This explains the excitation and intensification of plasma turbulence with an increase in its temperature. It follows from expression (20) that the boundary of excitation or attenuation of turbulence or "magnetic dynamo", in the non-relativistic approximation, is the equality $\nabla \mathbf{V}^2/2 = \mathbf{V}\cdot\nabla\cdot\mathbf{V}$.

Let's write Eq. (17) in the form:

$$\frac{1}{c^2}\partial_{tt}\mathbf{V} - \Delta\mathbf{V} - \nabla\times\nabla\times\mathbf{V} + \partial_t(\nabla\mathbf{V}^2/2 - \mathbf{V}\cdot\nabla\cdot\mathbf{V})/c^2 = 0 \qquad (20)$$

The first two terms describe classical charge velocity waves. Since the wave equation has nontrivial solutions when the wave equation is equal to zero, then for this particular case we equate the wave part to zero, assuming that $\mathbf{V} \neq 0$:

$$\frac{1}{c^2}\partial_{tt}\mathbf{V} - \Delta\mathbf{V} = 0 \qquad (21)$$

We take this wave part out of Eq. (20) and obtain a nonlinear equation:

$$\partial_t(\nabla \mathbf{V}^2/2 - \mathbf{V}\cdot\nabla\cdot\mathbf{V})/c^2 = \nabla\times\nabla\times\mathbf{V} \qquad (22)$$

This equation is no longer a wave equation. The movement of charges, determined by the right side, has a rotational toroidal character. This equation, depending on the sign of its left part, describes the excitation or attenuation of local rotating structures formed by moving charges. If the left part is constant, then we have a stationary rotational toroidal motion of charges. Since the toroidal movement of charges leads to the appearance of a poloidal magnetic field, these rotating structures must excite the corresponding stationary magnetic field and they can be associated with "ball lightning" and other local rotating plasma structures [12-14]. If the left side of equation (22) changes in time, then depending on its sign, this equation describes the excitation or attenuation of these rotating plasma structures. Such rotating structures can be formed in small limited volumes of plasma, where waves described by Eq. (21) cannot exist, this fact is noted in experimental work [13].

**Conclusion**

From Maxwell's equations with electromagnetic field sources, wave equations follow not only for the electromagnetic field, but also the wave equation for the density of the self-consistent current. This equation has the same degree of generality in describing the motion of charges as the equations of motion of the field. From the wave equation for the current density follows a nonlinear wave equation for the velocity of the charges and the equation of self-consistent rotating current structures of the plasma. The wave motion of free electric charges has a vortex component, which is a source of plasma turbulence. These equations provide a qualitative explanation of the mechanism of excitation of the "magnetic dynamo" and plasma turbulence and show that these phenomena express the basic properties of the motion of electric charges arising from Maxwell's equations without any additional conditions or assumptions. Maxwell's equations do not include the mass of electric charges, so it is also absent in the resulting equations, i.e. they describe the motion of massless charges. The introduction of the mass of charges into the equations, as is done in magnetic hydrodynamics, will not fundamentally change the electromagnetic laws that determine the self-consistent motion of charges, but only change the dynamics of their motion associated with the inertia of the mass.

**Appendix A**

Having solved the system of equations (14) – (15) with respect to $\partial_t\rho$ and $\nabla\rho$, we obtain the expressions:

$$\partial_t\rho = \rho \cdot (\frac{1}{c^2}\cdot\mathbf{V}\cdot\partial_t\mathbf{V} - \nabla\cdot\mathbf{V})/(1-\frac{V^2}{c^2}) \qquad (1)$$

$$\nabla\rho = \rho \cdot \frac{1}{c^2}\cdot(\mathbf{V}\cdot\nabla\cdot\mathbf{V} - \partial_t\mathbf{V})/(1-\frac{V^2}{c^2}) \qquad (2)$$

Taking the gradient from expression (1) and the time derivative of expression (2), we obtain the following expressions:

$$\nabla \partial_t \rho = \nabla(\rho \cdot (\frac{1}{c^2} \cdot \mathbf{V} \cdot \partial_t \mathbf{V} - \nabla \cdot \mathbf{V})/(1 - \frac{V^2}{c^2})) \quad \partial_t \nabla \rho = \partial_t(\rho \cdot \frac{1}{c^2} \cdot (\mathbf{V} \cdot \nabla \cdot \mathbf{V} - \partial_t \mathbf{V})/(1 - \frac{V^2}{c^2})) \quad (3)$$

Given the equality of the left parts of these expressions, we equate their right parts and get:

$$\nabla(\rho \cdot (\frac{1}{c^2} \cdot \mathbf{V} \cdot \partial_t \mathbf{V} - \nabla \cdot \mathbf{V})/(1 - \frac{V^2}{c^2})) = \partial_t(\rho \cdot \frac{1}{c^2} \cdot (\mathbf{V} \cdot \nabla \cdot \mathbf{V} - \partial_t \mathbf{V})/(1 - \frac{V^2}{c^2}) \quad (4)$$

By opening the brackets and performing differentiation, we obtain the equation:

$$M \cdot \nabla \rho + \rho \cdot \nabla M = \mathbf{N} \cdot \partial_t \rho + \rho \cdot \partial_t \mathbf{N} \quad (5)$$

Here $M = (\frac{1}{c^2} \cdot \mathbf{V} \cdot \partial_t \mathbf{V} - \nabla \cdot \mathbf{V})/(1 - \frac{V^2}{c^2}) \quad \mathbf{N} = \frac{1}{c^2} \cdot (\mathbf{V} \cdot \nabla \cdot \mathbf{V} - \partial_t \mathbf{V})/(1 - \frac{V^2}{c^2})$

Replacing in Eq. (5) $\partial_t \rho$ and $\nabla \rho$ by their expressions (1) and (2), we obtain:

$$M \cdot \rho \cdot \mathbf{N} + \rho \cdot \nabla M = \mathbf{N} \cdot \rho \cdot M + \rho \cdot \partial_t \mathbf{N} \quad (6)$$

By giving such terms and reducing p, we obtain a nonlinear vector kinematic equation for the velocity $\mathbf{V}$ of charges $\nabla M = \partial_t \mathbf{N}$, or in expanded form:

$$\nabla[(\frac{1}{c^2} \cdot \mathbf{V} \cdot \partial_t \mathbf{V} - \nabla \cdot \mathbf{V})/(1 - \frac{V^2}{c^2})] - \frac{1}{c^2} \cdot \partial_t[(\mathbf{V} \cdot \nabla \cdot \mathbf{V} - \partial_t \mathbf{V})/(1 - \frac{V^2}{c^2})] = 0 \quad (7)$$

**References**

1. Kocharovsky V V et al. Sov. Phys. Usp. **59** 12 (2016)
2. Moffatt H K. Magnetic field generation in electrically conducting fluids. Cambridge University Press, Cambridge. London-New York-Melbourne. (1978).
3. Vainshtein S I, Zel'dovich Ya B. Sov. Phys. Usp. **15** 159 (1972).
4. Sokolov D D et al. Sov. Phys. Usp. **57** 292 (2014)
5. Xu Siyao, Lazarian A. Small-scale turbulent dynamo in astrophysical environments: nonlinear dynamo and dynamo in a partially ionized plasma. arXiv:2106.12598v1 (2021).
6. Steinwandel U P et al. On the small scale turbulent dynamo in the intracluster medium: A comparison to dynamo theory. arXiv:2108.07822v1 (2021).
7. Singer S. The Nature of Ball Lightning. N.-Y. Plenum Press (1971).
8. Barry J D. Ball lightning and Bead Lightning, N.-Y. Plenum Press (1980).
9. Smirnov B M Physics Reports, 224 (1993) 151.
10. Funaro, D. Ball Lightning as Plasma Vortexes: A Reinforcement of the Conjecture. Appl. Sci. (2022), 12, 3451.
11. Neil J G. Retrospective Observation of Ball Lightning. arXiv:2204.08288v1 (2022).

12. Skvortsov V A, Vogel N I. The magnetic monopoles generation in laser-induced discharges. Proc. 29th EPS Conference on Plasma Physics and Contr. Fusion. Montreux, 17-21 June 2002 ECA, Vol.26 B. D-5.013 (2002).
13. Skvortsov V A, Vogel N I. The generation of exotic quasiparticles. In book: "Particle Physics in Laboratory, Space and Universe". World Scientific, New Jersey, London, Singapore, (2005).
14. Pirozerski A L et al. Artificial fireball generation via an erosive discharge with tin alloy electrodes. arXiv:1504.01120 (2015).
15. Novozhilov Yu V, Yappa Yu A. Elektrodinamika. M. Nauka. (1978).
16. Landau L D, Lifchits E M. Electrodynamics of Continuous Media. Vol. 8 (1st ed.). Pergamon Press. (1960).
17. Farlow S J. Partial differential Equations for Scientists and Engineers. John Wiley & Sons, Inc. (1982).